\documentstyle[12pt]{article}
\begin{document}
\title{\bf{Features of Single Lepton
Production at Hadron Collider}}\par

\author{I.Avaliani,\thanks{E-mail: IRA@BRE.GE} \\
{\sl{High Energy Physics Institute, Tbilisi State University}}\\
{\sl{380086 Tbilisi, Republic of Georgia}}}

\maketitle

\begin{abstract}
In the Born approximation, as in the next-to-leading
order approximation, Single lepton production $p\bar{p}\rightarrow l^{\pm}\nu
X$
at Fermilab Tevatron is discussed.
The effects of non standard model physics ($W'$ - production)
are also considered.
It  was shown  that
$2\rightarrow 3$ subprocesses play important role
when a high transverse momentum cut on the lepton is imposed.
On the other hand the NLO corrections do not cause qualitative changes of Born
approximation results. The contributions of $W'$ bosons for this signature
become more important for large values of $P_{T}^{cut}$.

\end{abstract}

\ \par
\ \par
Nearly all experimental data, which are available at present, support the
Standard Model as being the correct description of observables at current
energies. The signature of $W$ boson production at hadron colliders is
characterized by observing isolated leptons at a large transverse momentum
accompanied by a significant missing transverse momentum due to the
neutrino. The cross section for these processes is roughly a nanobarn. As
in the near future at the Fermilab Tevatron it will be possible to measure
these cross sections of the order 0.1pb, there is a chance to find the
effects beyond the Standard Model. To be sure that these effect are the
results of the new physics, one needs more precise calculations in the frame
of the Standard Model (like higher order corrections and etc).

One of the possibilities of the new physics is the existence of new
neutral or charged vector bosons, what  is a common feature of many
extensions of the Standard Model.
Different models give different predictions for the production of new
heavy vector bosons\cite{1}-\cite{10}.~ Below we consider a simple model
obtained by taking
for the heavy $W'^\pm$ and $Z'$ gauge bosons the same coupling to fermions
as ordinary $W^\pm$ and $Z$\cite{W'}. In this model there exist also the
trilinear couplings $W'WZ$ and $Z'WW$, which give the important contributions
to the decay widths of $W'$ and $Z'$ bosons.

We shall consider events with isolated leptons in the final state with a large
transverse momentum - $p\bar{p}\rightarrow l^{\pm}\nu X$. A possible
contribution to this signature from the new physics can be for example
$W'$ boson production\cite{W'}. So it is important to know whether the cross
section
is fully determined by the Standard Model or not.

In the lowest order the cross section of the reaction  $p\bar{p}\rightarrow
l^{\pm}\nu X$
is determined by the subprocess
\begin{equation}
q\bar{q'}\rightarrow W^\pm \rightarrow l^{\pm}\nu  \label{enu}
\end{equation}
On the other hand it is also necessary to consider the subprocesses:
\begin{equation}
q\bar{q'}\rightarrow W^{\pm}g\rightarrow l^{\pm}\nu g, ~~~~~
qg\rightarrow W^{\pm}q\rightarrow l^{\pm}\nu q   \label{enug}
\end{equation}
If the kinematical parameters (the transverse momentum $p_T$ or  the rapidity
$|\eta |$)
of a gluon or a quark in the final state do not satisfy the triggers
requirements
they can be not detected. So there are  kinematical regions, where (\ref{enu})
and (\ref{enug}) subprocesses give the similar signatures. Therefore it is
important to calculate the contribution of subprocesses (\ref{enug}) to the
cross section of $p\bar{p}\rightarrow l^{\pm}\nu X$.

Taking into account subprocesses (\ref{enu}), (\ref{enug}),
below we consider the $p\bar{p}\rightarrow l^{\pm}\nu X$ cross section.
For calculation of this cross section we use the well known in QCD expression
of $ \sigma (AB\rightarrow CX)$, which after making the following change of
variables
$$
w=x_a x_b, \; \; \; z=x_a/x_b
$$
we can rewrite in more convenient for us form:
\begin{equation}
\sigma(AB\rightarrow C_1
X)=\frac{1}{2}\sum_{a,b}\int_{w_0}^{1}dw\int_{z_-}^{z_+} \frac{dz}{z}
f_{a/A}(\sqrt{w z},Q^2) f_{b/B}(\sqrt{\frac{w}{z}},Q^2) \hat{\sigma}(a
b\rightarrow c_1+\cdots )  \label{sig}
\end{equation}
where the sum is taken over all the subprocesses which lead to the
production of particles $C_1,\ldots$ in the final state. $\hat{\sigma}$ is the
total cross section for subprocess $a+b\rightarrow c_1+\cdots$ and $f_{h/H}$ is
a distribution function of parton $h$ in a hadron $H$.
$z_{-}=w, z_{+}=1/w$ and $w_{0}=\hat{s}_{min}/S$ ($\hat{s}_{min}$
is the threshold value of partonic squared center of mass energy $\hat{s}=w S$
for
the reaction of interest).

The total cross section $\hat{\sigma}$ of $2 \rightarrow 3$ subprocess can be
written in a
standard form \cite{byckling}
\begin{equation}
\hat{\sigma}(p_a+p_b\rightarrow
p_1+p_2+p_3)=\frac{1}{2(2\pi)^5\hat{s}}\frac{\pi}{16\hat{s}}\int_{\hat{t}_1^-}^{\hat{t}_1^+}d\hat{t}_1\int_{\hat{s}_2^-}^{\hat{s}_2^+}d\hat{s}_2\int_{\hat{t}_2^-}^{\hat{t}_2^+}d\hat{t}_2   \label{sig23}
\int_{\hat{s}_1^-}^{\hat{s}_1^+}d\hat{s}_1\frac{|\overline{M}|^2}{\sqrt{-\Delta_4}}
\end{equation}
where $|\bar{M}|^2$ is a squared matrix element and $\Delta_4$ is Gram
determinant
of the fourth order.

The selection procedure of events used in experiments at hadron colliders
requires certain experimental cut-offs of kinematical parameters. As a rule
the finite resolution of a detector requires the cut-off of a transverse
momentum of jets(leptons)
$$
 |\vec{p}_T| \geq |\vec{p}_T^{min}|\equiv q_T
$$
and their polar angles (pseudorapidity)
$$
\pi - \theta_0 \geq \theta\geq \theta_0
$$
$p_T$ and $\theta$ are assumed to measure in the c.m.s. of the colliding
hadrons.

If we  consider the case when one of the
jets'(leptons') transverse momentum $p_T$ and polar angle $\theta$ satisfy the
above requirements and ignore all masses of particles(partons)
in the initial and final states, then $\hat{s}_{min}= 4q_T^2$ and consequently
$w_0=4q_T^2/S$. From the relations
$(sin\hat{\theta})_{min}={2q_T}/{\sqrt{\hat s}}={2q_T}/{\sqrt{w S}}$,
$(sin\theta)_{min}={2q_T}/{\sqrt{S}}$ ($\hat{\theta}$ is a polar angle in the
c.m.s.
of partons) we can determine the values of  $z_{+}, z_{-}$
in the case of kinematical cut-offs $|\vec{p}_T| \geq |\vec{p}_T^{min}|\equiv
q_T ,
\pi - \theta_0 \geq \theta\geq \theta_0$:
\begin{equation}
z_{+} = min \left [\frac{1}{\omega},\frac{(1+a)(1+A)}{(1-a)(1-A)}\right ],~~~~~
z_{-} = max \left [\omega,\frac{(1-a)(1-A)}{(1+a)(1+A)} \right ]  \label{z+-}
\end{equation}
where $a\equiv \sqrt{1-\frac{4q_T^2}{wS}},  A\equiv
min(\sqrt{1-\frac{4q_T^2}{S}}, \cos\theta_0)$.
When $\theta_{0}, \rightarrow 0$ $z_{+}, z_{-}$ transfer into the standard
expressions $z_{-}=w, z_{+}=1/w$ for the three body phase space.

Very often in experiments we deal with the case when the transverse momentum
and polar angle of one jet (lepton) are above $q_T$ and $\theta_0$, but the
transverse momentum of another jet (lepton) is below $k_T$ -
$|\vec{p}_T| \leq k_T$. It should be noted that the expressions (\ref{z+-})
are valid for this case too.

The kinematical cut-offs of $p_{1T}$ and $\theta$ (for the particle
in the final state with the momentum $p_1$) define  the bounds of
integrals in (\ref{sig23}).
By using the expressions for $p_{1T}$ and $\tan(\theta/2)$,
\begin{equation}
p_{1T}=\sqrt{-\frac{\hat{t}_1(\hat{t}_1+\hat{s}-\hat{s}_2)}{\hat{s}}}; \; \; \;
 \tan(\theta/2)=\sqrt{-\frac{\hat{t}_1}{z(\hat{t}_1+\hat{s}-\hat{s}_2)}}
\label{pt}
\end{equation}
and after taking into account the kinematical cut-offs $\vec{p}_{1T}\geq q_T,
\theta_0 \leq \theta_1 \leq \pi-\theta_0$ we get:
\begin{equation}
\hat{s}_1^-=\max[0,\hat{t}_2-\hat{t}_1],~~~~
\hat{s}_1^+=\min[\hat{s}-\hat{s}_2, \hat{s}+\hat{t}_2]  \label{s1}
\end{equation}

\begin{equation}
\hat{t}_2^-=\hat{t}_1-\hat{s}_2,~~~~  \hat{t}_2^+=0  \label{t2}
\end{equation}

$$
\hat{s}_2^-=max \left [0,\hat{t}_1+\hat{s}+\frac{\hat{t}_1}{z
\tan^2(\theta_0/2)}\right ]; \  \  \
\hat{s}_2^+=min \left [\hat{t}_1+\hat{s}+\frac{q_T^2\hat{s}}{\hat{t}_1},
\hat{t}_1+\hat{s}+\frac{\hat{t}_1\tan^2(\theta_0/2)}{z}\right ]
$$

\begin{eqnarray}
\hat{t}_1^+=min\left
[-\frac{2q_T\sqrt{\hat{s}}}{1+\frac{1}{z\tan^2(\theta_0/2)}};
-\frac{\hat{s}}{2}+\frac{\hat{s}}{2}\sqrt{1-\frac{4q_T^2}{\hat{s}}}\right ]
\nonumber \\
\hat{t}_1^-=max\left [-\frac{\hat{s}}{1+\frac{\tan^2(\theta_0/2)}{z}};
-\frac{\hat{s}}{2}-\frac{\hat{s}}{2}\sqrt{1-\frac{4q_T^2}{\hat{s}}}\right ]
\end{eqnarray}
As it was mentioned above, because of a finite resolution of a detector
used in experiments at hadron colliders, only particles with $\vec{p}_T\geq
q_T$
and $\theta \geq \theta_0$ can be detected. So it can  happen that from
three partons' in a final state, produced via $2\rightarrow 3$ subprocess,
only two (for which $\vec{p}_T\geq q_T, \theta \geq \theta_0$) are detected.
To distinguish this signature from the other one, produced via ordinary
$2\rightarrow 2$
subprocess, we need additional cut-off of a transverse momentum of the third
particle
(jet or lepton) - $|\vec{p}_{3T}|\leq k_T$. In this case according to the
expression
$$
p_{3T}=\sqrt{-\frac{\hat{t}_2(\hat{t}_2+\hat{s}-\hat{s}_1)}{\hat{s}}}
$$
the values of bounds $\hat{s}_1^-,\hat{s}_1^+$ must be changed:

\begin{equation}
\hat{s}_1^-=\max\left [0, \hat{t}_2-\hat{t}_1,
\hat{s}+\hat{t}_2+\frac{\hat{s}k_T^2}{\hat{t}_2} \right ]; ~~~~
\hat{s}_1^+=\min \left [\hat{s}-\hat{s}_2, \hat{s}+\hat{t}_2\right ]
\label{s11}
\end{equation}

The corresponding squared amplitudes of subprocesses (\ref{enu}) and
(\ref{enug}),
summed over final state spins and colors
and averaged over initial spins and colors have the following form:
\begin{eqnarray}
\frac{d\hat{\sigma}}{d\hat{t}}(q\bar{q'}\rightarrow l^{+}\nu) = \frac{\pi
\alpha^{2} |V_{qq'}|^{2}}{12{\hat{s}}^2 \sin^{4}\theta_W}
\frac{\hat{t}^2}{[(\hat{s}-m^{2}_W)^{2} + m_{W}^2 {\Gamma}_{W}^2]} \\
|\bar{M}(q\bar{q'}\rightarrow l^{+}\nu g)|^2 = \frac{2 (4\pi)^3 \alpha_S
\alpha^{2} |V_{qq'}|^{2}}{9 \sin^{4}\theta_W} \frac{\hat{s}_1
[\hat{t}_{1}^2+(\hat{s}_2+\hat{t}_2-\hat{t}_1)^{2}]}{(-\hat{t}_2)(\hat{s}-\hat{s}_1+\hat{t}_2)
[(\hat{s}_{1}-m^{2}_W)^{2} + m_{W}^2 {\Gamma}_{W}^2]} \\
|\bar{M}(qg \rightarrow l^{+}\nu q')|^2 = \frac{(4\pi)^3 \alpha_S \alpha^{2}
|V_{qq'}|^{2}}{12\sin^{4}\theta_W} \frac{\hat{s}_1
(\hat{t}_{1}^2+\hat{s}_{2}^2)}{\hat{s}(-\hat{t}_2)[(\hat{s}_{1}-m^{2}_W)^{2} +
m_{W}^2 {\Gamma}_{W}^2]}
\end{eqnarray}
where $\alpha_S$ is the strong coupling constant, $\alpha$ - the fine structure
constant, $V_{qq'}$ - the Cabibbo-Kobayashi-Maskava mixing matrix, $m_W,
\Gamma_W$
- the mass and the total width of $W$ boson. Our results for $2\rightarrow 3$
squared amplitudes coincide with the corresponding expressions of \cite{baer},
but are written in a more convenient form.\par

In our computations we choose  $Q^2 = \hat{s}$ and we take the structure
functions from \cite{duke}. For the events which could represent the signals
of the new physics - $p_{T}^l \geq m_W$, the area of relatively
large $x (x \geq 0.03)$ is the most important. Therefore  different choices for
the structure
functions do not make a significant difference for this area. As at CDF
additional cuts are
used to reject hadronic clusters with transverse energy $E_T \geq 7GeV$, we
take $k_T = 7GeV/c$. Therefore it is required that the transverse
momenta of partons $g$ and $q$ in the final state of subprocesses (\ref{enug})
are less than $7GeV/c (|\vec{p}_{3T}|\leq k_T = 7GeV/c)$.

In Fig.1 the total cross section of $p\bar{p}\rightarrow l^{+}\nu X$ reaction
($\sqrt{S}=1.8TeV$) is presented. The dotted line corresponds to the
production of a single species of lepton with the transverse momentum $p_T\geq
p_{T}^{cut}$
and $|\eta^l| \leq 2$, generated by $2\rightarrow 2$ subprocess (\ref{enu}),
while the dashed line corresponds to the contribution generated by
$2\rightarrow 3$
subprocesses (\ref{enug}). For this last case it is assumed that there are no
hadronic
clusters (generated by outgoing partons) with $E_T\geq 7GeV.$
The solid line represents their sum - the total
cross section. It is evident that for the values of the transverse momentum
$p_{T}^{cut}\geq 50GeV/c$ the cross section with $2\rightarrow 3$  subprocesses
will prevail over the ordinary cross section generated by $2\rightarrow 2$
subprocess. The ratio $\sigma(2\rightarrow 3)/\sigma(2\rightarrow 2)$ for
large values of $p_{T}^{cut}$ reaches to 3 and therefore the role of higher
order contributions to the cross section become important while making a
decision about the role of new physics in this area.

In the case of existence of $W'$ boson there are analogous to (\ref{enu}),
(\ref{enug}) contributions to the cross section of the reaction
$p\bar{p}\rightarrow l^{+}\nu X$. In fig.2 the total cross sections of this
reaction generated by $W'$ bosons of mass $m_{W'}=200GeV/c$ are presented.
The dotted line corresponds to $2\rightarrow 2$ subprocess (\ref{enu}),
while the dashed line to the sum of $2\rightarrow 3$ subprocesses (\ref{enug}).
The solid line represents their sum. All parameters are the same as in fig.1:
$|\eta^l| \leq 2$ and $k_T = 7GeV/c$.

In fig.3 we plot the values of the total cross section $p\bar{p}\rightarrow
l^{+}\nu X$
in the case - of the Standard Model and of the existence of $W'$ bosons
\cite{W'}.
Here the sum over the $2\rightarrow 2$ and $2\rightarrow 3$ subprocesses is
assumed and $|\eta^l| \leq 2$, $k_T = 7GeV/c$. It is clear that according
to the model \cite{W'} the contributions of $W'$ bosons for this signature
become important for large values of $p_T^{cut}$. It proves once again the
necessity of more precise estimation of this cross section in the Standard
Model.

For more precision, when considering $O(\alpha_s)$ corrections to the
lowest-order diagrams, besides the real emission subprocesses (\ref{enug}),
one has to include the interference between the loop corrections to
$q\bar{q'} \rightarrow \bar{l} \nu $ and the Born graphs.
There are several computations on higher order corrections to $W$ boson
production including also the decays \cite{ALT},\cite{au},\cite{ar}.
For calculation of
soft collinear and virtual singularities we use the method of \cite{baer},
dividing the phase space into singular and nonsingular regions. According to
\cite{baer} we introduce two cutoffs $\delta_s$ and $\delta_c$. When
\begin{equation}
E_g < \delta_s \sqrt{\hat{s}}/2   \label{eg}
\end{equation}
the soft-gluon approximation is used to evaluate $q\bar{q'} \rightarrow
g\bar{l}\nu$
subprocess and when
\begin{equation}
|\hat{t}| < \delta_c \hat{s}   \label{tc}
\end{equation}
the diagrams are evaluated in the leading-pole approximation.

Again the cross section of the reaction $p\bar{p} \rightarrow l^{\pm} \nu X$
is determined by two contributions
$$
\sigma (p\bar{p} \rightarrow l^{\pm} \nu X) = \sigma^{NLO}_{2\rightarrow 2} +
\sigma_{2\rightarrow 3}
$$
where $ \sigma^{NLO}_{2\rightarrow 2}$ according to \cite{baer} has the
following
form (we assume that $\lambda_{FC} = 1$ and $\hat{s}=Q^2\equiv M^2$ ):
\begin{equation}
\sigma^{NLO}_{2\rightarrow 2} = \sigma^{HC} + \sum_{q_1,\bar{q'_2}}\int dx_a
dx_b f_{q_{1}/A}(x_a,Q^2) f_{\bar{q'_2}/B}(x_b,Q^2) \hat{\sigma}^{NLO}
\label{nlo}
\end{equation}
Here $\sigma^{HC}$ is the contribution from the hard collinear remnants, and
\begin{equation}
\hat{\sigma}^{NLO} = \hat{\sigma}^{Born}[1+\frac{\alpha_s}{2\pi}\frac{4}{3}
[1+5\pi^2/3 + 2{\ln{(\delta_s)}}^2 + 3\ln{(\delta_s)}]]
\end{equation}

There are two contributions from the remnants of the hard collinear
singularities $\sigma^{HC} = \sigma^{HC}_{q\bar{q'}} + \sigma^{HC}_{qg}$
\begin{eqnarray}
\sigma^{HC}_{q\bar{q'}} =
\sum_{q_1,\bar{q'_2}}\int dx_a dx_b \frac{\alpha_s}{2\pi}
\int_{x_b}^{1-\delta_s} \frac{dz}{z}
 f_{q_{1}/A}(x_a,Q^2) f_{\bar{q'_2}/B}(\frac{x_b}{z},Q^2) \hat{\sigma}^{Born} *
 \nonumber \\
 \frac{4}{3} [\frac{1+z^2}{1-z^2}\ln(\delta_c)+\frac{3}{2}\frac{1}{1-z}-2-3z]
+(x_a \leftrightarrow x_b)
\end{eqnarray}
and
\begin{eqnarray}
\sigma^{HC}_{qg} =
\sum_{q_1,\bar{q'_2}}\int dx_a dx_b \frac{\alpha_s}{2\pi} \int_{x_b}^{1}
\frac{dz}{z}
 f_{q_{1}/A}(x_a,Q^2) f_{g/B}(\frac{x_b}{z},Q^2) \hat{\sigma}^{Born} *
\nonumber \\
 \frac{1}{2} [(z^2 + (1-z)^2)\ln(\delta_c) + 1 - 6z(1-z)] +(x_a \leftrightarrow
x_b)
\end{eqnarray}

In the case of the next-to-leading order calculations, in principle, we should
use the NLO distribution functions, however below we use the LL distributions
\cite{duke}.

For calculation of the $\sigma_{2\rightarrow 3}$ cross section we have to
consider that
\begin{eqnarray}
E_g > \delta_s \sqrt{\hat{s}}/2   \\
|\hat{t_2}| > \delta_c \hat{s}
\end{eqnarray}
which leads to the following changes in equations (\ref{t2}) and (\ref{s11})
\begin{eqnarray}
t_2^{+} = -\delta_c \hat{s}  \\
\hat{s}_1^+=\min[\hat{s}-\hat{s}_2, \hat{s}+\hat{t}_2, \hat{s}(1-\delta_s)]
\end{eqnarray}
$\delta_s$ and $\delta_c$ are not the physical parameters. So the inclusive
cross section must not depend on them, or at the worst, must depend on them
weakly.
The analytical and numerical analyses of phase space show that for
experimental cuts used at CDF (considered above) the total cross section
$\sigma_{2\rightarrow 3}$
does not depend on $\delta_s $in the case when $\delta_s < 0.05$. (large values
of $\delta_s$ and $\delta_c$ are not desirable as we use the soft gluon and the
leading-pole
approximations while evaluating the diagrams). So, the dependence on $\delta_s$
of inclusive cross section is due to $\hat{\sigma}^{NLO}$ and
$\sigma_{q\bar{q'}}^{HC}$
terms. As the dependence of the cross section on $\delta_s$ must be weak, we
use the equation
$$
\frac{\partial\sigma_{2\rightarrow 2}^{NLO}}{\partial\delta_s} = 0
$$
to determine the corresponding values of $\delta_s$. We get
\begin{equation}
\delta_c \approx \exp{(3/4)} \delta^{2}_s \label{dc}
\end{equation}
This relation practically does not depend on the shape of the distribution
functions.

On the other hand, according to the numerical calculations of $p\bar{p}
\rightarrow l^{\pm} \nu X$,
the values of $\delta_c$, determined as the solutions of the equation
$$
\frac{\partial\sigma(p\bar{p} \rightarrow l^{\pm} \nu X)}{\partial\delta_c} = 0
$$
are not constant and depend weakly on the values of $P_T^{cut}\equiv q_T$.

In the case of experimental cuts and distribution functions considered above,
the values of $\delta_c$ vary in the interval 0.002 - 0.004 for
$20GeV < P_T^{cut} < 160 GeV$. If we consider that approximately $\delta_c =
0.003$,
than according to (\ref{dc}) $\delta_s \approx 0.04$. We use these values to
calculate
the total cross section $\sigma(p\bar{p} \rightarrow l^{\pm} \nu X)$ in the NLO
approximation. We present our result in fig.4. The solid line represents the
NLO approximation, while the dashed line - the Born contributions
($2\rightarrow 2$
and $2\rightarrow 3$ subprocesses). The calculation shows that the ratio
$\sigma^{NLO} / \sigma^{Born}$ for the given interval of $P_{T}^{cut}$ varies
in
the range 0.7 - 1.7. So these corrections do not cause qualitative changes of
$\sigma^{Born}$
results, what was mentioned also in \cite{baer}.

therefore we can conclude that
$2\rightarrow 3$ subprocesses play important role
when a high transverse momentum cut on the lepton is imposed.
On the other hand the NLO corrections do not cause qualitative changes of Born
approximation results. The contributions of $W'$ bosons for this signature
become more important for large values of $P_{T}^{cut}$.

We are grateful to V.Kartvelishvili and
M.Margvelashvili for fruitful discussions.

\ \par
\ \par
\par

\newpage
\begin{center}
{\Large{\bf{Figure Captions}}}
\end{center}\par
\ \par

{\bf Fig. 1}~ {The total cross section for events containing a single species
of
lepton with transverse momentum $p_{T}^l\ge P_{T}^cut$
and $|\eta^{l}|\leq 2$.  }

\ \par
{\bf Fig. 2}~ {The total cross section for events containing a single species
of lepton
generated by $W'$ bosons\cite{W'}  }

\ \par
{\bf Fig. 3}~ {The total cross section for events containing a single species
of lepton
generated by the Standard Model and $W'$ bosons\cite{W'} }

\ \par
{\bf Fig. 4}~ {The total cross section for events containing a single species
of lepton
in the Born and next-to-leading order approximations }

\end{document}